# Human Trajectories Characteristics

Suhad Faisal Behadili[1], Cyrille Bertelle[1], and Loay E. George[2]

[1]Normandie University, LITIS, FR CNRS 3638, ISCN, ULH, Le Havre, France
[2] Baghdad University, Computer Science Department, Baghdad, Iraq

*Abstract* : *Communication devices (mobile networks, social media platforms) are produced digital traces for their users either voluntarily or not. This type of collective data can give powerful indications on their effect on urban systems design and development. For understanding the collective human behavior of urban city, the modeling techniques could be used. In this study the most important feature of human mobility is considered, which is the radius of gyration $R_g$. This parameter is used to measure how (far /frequent) the individuals are shift inside specific observed region.*

## 1. Introduction

The population urban analysis is a wide range area has a multi-disciplinary subject to obtain density, community detection, crowd movements, and spatial coordination, which is the 21st century focuses, since it guides to a lot of sciences like: cities planning, smart cities, decision making, facing disasters, social networks analysis, diseases spreading, ecosystems evolution, human attitudes, human body inter-relations, networks understanding, society cultural analysis... etc. [11, 2, 3, 4, 5, 6, 7]. The simulation is preferred to analyze any system instead of working on it in real life, since it is easier, safer, cheaper to make all the experiments on the simulated model, the most important and difficult issue in modeling and simulating any system is the determination of the probability distribution and parameters to model the uncertainty of the system input variables. However, the purpose of simulation analysis is to acquire and analyze the results in well conceptual vision, in order to give high indications for decision makers, pivoting on two events kinds: simulation using discrete events, and simulation using frequent (continuous) events. The $R_g$ is used to describe the individuals trajectories during observation period, which is indeed the standard deviation of individual's positions to the center of mass [8, 9].

## 2. Data Set

The available mobile phone data of Armada case study composed of 51,958,652 CDRs (call detailed records) for 615,712 subscribers. However, it contains individuals occurrence in discrete (irrelevant) mode only, means that any mobile individuals' activity is recorded at (start/end) time, but there is a lack (lost information), which is supposed to indicate the user's occurrence during inactive case (mobility without any mobile phone activity). With regarding to the non-deterministic & discrete nature of this data, therefore the collective behavior would be the effective approach to be analyzed and simulated. Since each individual could be disappeared for a while from the DB records, which makes individual tracing is unworthy, without significant indications on the people behavior in the city [6]. However, the data of one day is elected from the whole data to be studied and elaborated, composed of 5,661,428 CDRs and 235, 400 subscribers.

## 3. Trajectories in Intrinsic Reference Frame

The significant importance of revealing the human trajectories enforces the tendency to build the statistical models. Human trajectories have random statistical patterns; hence tracing human daily activities is most challengeable issue, in addition to its importance which is mentioned earlier as urban planning, spread epidemics...etc. In spite of data sources variance (billing system, GSM, GPS), but the common characteristics are the aggregated jump size (Δr), and waiting time (Δt) distributions. Where, (Δr) gives an indication on the covered distances by an individual in (Δt) for each two consecutive activities, and the (Δt) is the time spent by an individual between each two consecutive activities [10].

The individual trajectory is considered as microscopic level of mobility abstraction, which is constituted of sequenced coordinates positions along time i.e. the agent displacement in spatio-temporal unit.

### 3.1. Individual Trajectory Characteristics

The CDRs real data of each individual has many lack of data followed a bursty pattern, since the recorded data are not continuous, instead it is discrete (locations irrelevant), because it is limited by individual's activities (just when any cellular activities are made), so when individual's cellular is inactive there is no data, hence this would produce irregular activities pattern. However, this lack of data needs some estimation to compensate them. The individual trajectory has time-invariant properties; however trajectories estimations need to be computed according to common individual mobility characteristics [1, 12, 13, 14, 15, 6, 16, 18]. The individual's activities sparseness causes incomplete spatial information, therefore the mobile individual has some general physical characteristics, that are useful to compensate the lack of data (when no activity recorded), that could be used to build mathematical model of human mobility patterns, in order to verify the behavior and life patterns using the most common mobility characteristics, which are as follows [30, 19, 20, 21, 22, 23, 6, 16, 24, 18]:

1. Center of mass: It's the most visited positions by individual $c_m$, as in the following equations (1-2):

$$x_{cm} = \sum_{i=1}^{n} x_i/n \qquad (1)$$
$$y_{cm} = \sum_{i=1}^{n} y_i/n \qquad (2)$$

Where $x_i$ and $y_i$ are the coordinates of the spatial positions, n is the number of spatial positions, that are recorded in the CDRs.

2. Radius of Gyration: It is the average of all individual's positions, which is the indication to the area visited by the individual (the traveled distance during time period), as formulated in equation (3), the distribution of $r_g$ uncovers the population heterogeneity, where individuals traveled in P(r) in (long/short) distances regularly within $r_g(t)$ the distribution p($r_g$) produces power law as in equation 13 investigated in the aggregated traveled distance distribution p($r_g$) as in figure 1.

$$r_g^a = \sqrt{\frac{1}{n_c^a(t)} \sum_{i=1}^{n_c^a} (\vec{r_i^a} - \vec{r_{cm}^a})^2} \qquad (3)$$

Where $\vec{r_i^a}$ refers to i=1... $n_c^a(t)$ positions recorded for individual a, and $\vec{r_{cm}^a} = \frac{1}{n_c^a(t)} \sum_{i=1}^{n_c^a} \vec{r_i^a}$, which refers to center of mass of the individual's trajectory.

3. Most frequent positions: To uncover the individual tendency according to the frequent visited locations.
4. Principal axes θ (moment of inertia): This technique makes it possible to study some individuals' trajectories in common reference frame, by diagonalizing each of trajectory inertia tensor, hence to compare their different trajectories. Moment of inertia to any object is obtained from the average spread of an object's mass from a given axis. This could be elaborated using two dimensional matrix called tensor of inertia, then by using standard physical formula the inertia tensor of individual's trajectories could be obtained, as in the following equations (4-9):

$$I = \begin{pmatrix} I_{xx} & Ixy \\ Iyx & Iyy \end{pmatrix} \qquad (4)$$

$$I_{xx} = \sum_{i=1}^{n} y_i^2 \qquad (5)$$

$$I_{yy} = \sum_{i=1}^{n} x_i^2 \tag{6}$$

$$I_{xy} = I_{yx} = -\sum_{i=1}^{n} x_i y_i \tag{7}$$

$$\mu = \sqrt{4I_{xy}I_{yx} + I_{xx}^2 - 2I_{xx}I_{yy} + I_{yy}^2} \tag{8}$$

$$\cos\theta = -I_{xy}(1/2I_{xx} - 1/2I_{yy} + 1/2I_\mu)^{-1} \frac{1}{\sqrt{1 + \frac{I_{xy}^2}{(1/2I_{xx} - 1/2I_{yy} + 1/2I_\mu)^2}}} \tag{9}$$

5. Standard Deviation: To verify the horizontal and vertical coordinates of individual mobility in the intrinsic reference frame. However, trajectories are scaled on intrinsic axes using standard deviation of the locations for each individual a as in equations (15-16).

$$\sigma_x^a = \sqrt{\frac{1}{n_c^a} \sum_{i=1}^{n_c^a}(x_i^a - x_{cm}^a)^2} \tag{10}$$

$$\sigma_y^a = \sqrt{\frac{1}{n_c^a} \sum_{i=1}^{n_c^a}(y_i^a - y_{cm}^a)^2} \tag{11}$$

Then, obtain universal density function as in equation 11:

$$\tilde{\phi} = (x/\sigma_x, y/\sigma_y) \tag{12}$$

However, using the spatial density function to aggregate the individuals with the common $r_g$'s

The $R_g$ Probability distribution is computed for the individuals' population with respect to the data time sorting with power law of equation 13:

$$P(r_g) = (r_g)^{\exp(-r_g)} \tag{13}$$

As in figures (1-4), in order to classify them according to their $(R_{g_s})$, then a group of individuals had been chosen according to their pertinence of $(R_g)$, where each one is elected randomly from different $(R_g)$ sample, then the potential trajectories of the individuals $(user_1)$, $(user_2)$, $(user_3)$ are computed and simulated as in the figure 5. The $(R_g)$ probability distribution reveals that $R_g$ will reaches the stability, when it is increased, and the observed $R_g$ is small according to the limited studied region under observation. As well as, most of individuals are mobile in $R_g < 10$ Km, while in second level the others which are mobile in $10 < R_g < 20$ Km, however in the lowest are the individuals which mobile in $R_g > 20$.

The trajectories would be rescaled for group of individuals according to the mobility characteristics mentioned earlier, by choosing individuals from different classified groups with regarding to their Radius of Gyration, as in figure 2:

## 4. Conclusions

The simulation process represents the characteristics and behavior of any system. The word (simulation) is used in several fields, which includes the modeling of natural sentences or human organs as a try to explore the details of this process. There is also a simulation in technical and safety engineering, where the objective examination of some of the work in the real world scenarios, and test the security of some operations or any scientific and economic feasibility. Data analysis could be done in two levels [25, 26, 27]:
1. Individual level: The data of space, time or both of them, points to individual items.

2. Collective (group) level: The data of aggregated space or aggregated time intervals or both of them, point to aggregated data items.

Computing the p ($r_g$) is done for all users to approximate the mobility patterns, which is followed power-law. This law gives the indication of population heterogeneity, which characterized the patterns of identical individuals' trajectories. It is clear that there is a variance in individual mobility patterns $r_g$, as well as the individuals' trajectories are bounded by their $r_g$. The radius of gyration is the most common quantity, which is associated with human mobility trajectories, due to its capability in measuring the how far the mass from center of mass. The $r_g$ gradually increases at the beginning, but it settles down versus time. It has key effect on the travel distance distributions. The traveled distance distributions are collapsed or overlapped for groups, however the $r_g$ is considered to be more dedicated feature that capable of characterizing the travel distances Δr of individuals. The distributions showed that the individuals travel activities are almost identical, here the periodic trajectories are invariant, as well as the sedentary individuals have small $r_g$, in opposite to mobile individuals which have large $r_g$. Also, the activities distributions are uncovered the regular patterns and behaviors similarities, during the time evolution of $r_g$. The experiments analyzed the relationships between the $r_g$, Δr and the $r_{cm}$, and shows that all individuals have almost similar activities.

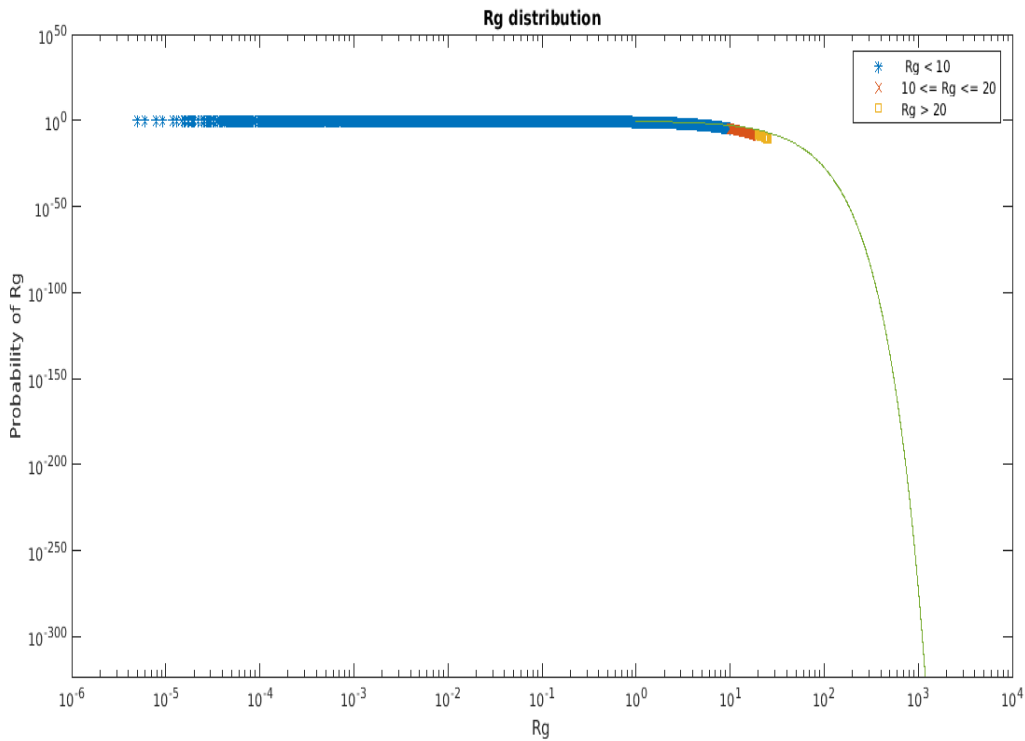

Fig. 1: Probability Distribution for the Radius of Gyration

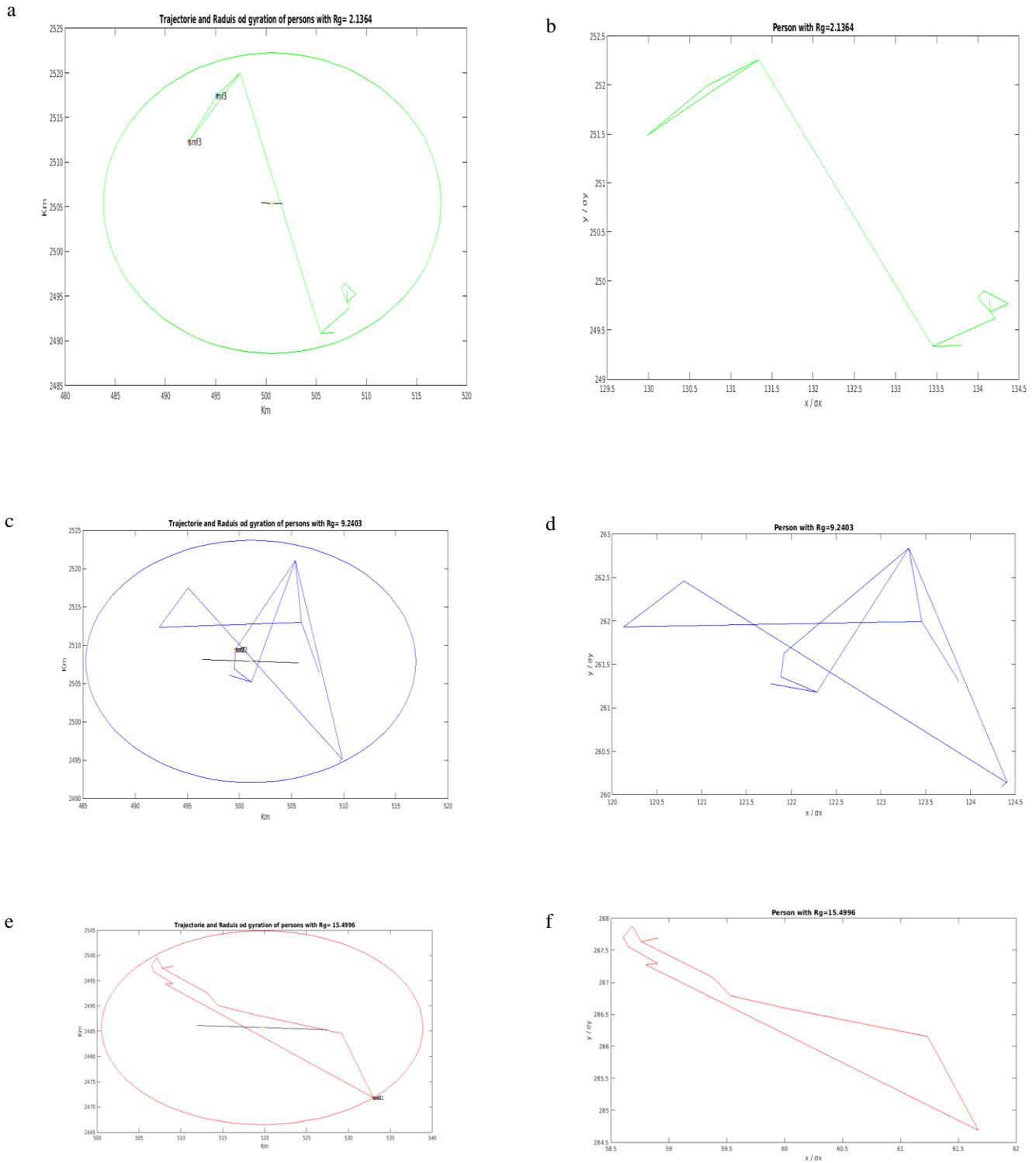

Fig. 2: Estimated Individuals' Trajectories within Intrinsic Reference Frame: a. Individual Trajectory Chosen from $R_{g2}$ with colored circles (red/blue) point to most/second most frequent positions, b. Individual Scaled Trajectory Chosen from $R_{g2}$, C. Individual Trajectory Chosen from $R_{g9}$ with colored circles (red/blue) point to most/second most frequent positions, d. Individual Scaled Trajectory Chosen from $R_{g9}$, e. Individual Trajectory Chosen from $R_{g15}$ with colored circles (red/blue) point to most/second most frequent positions, f. Individual Scaled Trajectory Chosen from $R_{g15}$.